\def\ask{\langle S_k \rangle}
\def\ak{\langle k \rangle}
\def\om{\Omega}
\def\ota{one-to-all }

\def\tt{\tilde t}

\newcommand{\xo}[1]{x_{[1] #1}}
\newcommand{\Xo}[1]{\tilde{X}_{[1] #1}}
\newcommand{\xt}[1]{x_{[2] #1}}
\newcommand{\Xt}[1]{\tilde{X}_{[2] #1}}
\newcommand{\po}[1]{p_{[1] #1}}
\newcommand{\Po}[1]{\tilde{P}_{[1] #1}}
\newcommand{\ptw}[1]{p_{[2] #1}}
\newcommand{\Pt}[1]{\tilde{P}_{[2] #1}}
\newcommand{\Mo}[1]{\tilde{M}_{ #1}}
\newcommand{\Mt}[1]{\tilde{M}_{ #1}}
\newcommand{\Vo}[2]{V_{ #1 #2}\left( \xo{#1}-\xo{#2} \right)}
\newcommand{\Vt}[2]{V_{ #1 #2}\left( \xt{#1}-\xt{#2} \right)}

\def\be{\begin{equation}}
\def\ee{\end{equation}}
\def\te{\end{equation}}
\def\bea{\begin{eqnarray}}
\def\eea{\end{eqnarray}}
\def\ea{\end{eqnarray}}
\def\tea{\end{eqnarray}}

\def\(#1){(\ref{#1})}

\newskip\humongous \humongous=0pt plus 1000pt minus 1000pt

\newif\ifdtup

\documentclass[a4paper]{jpconf}
\usepackage{lmodern}
\usepackage{amsfonts}
\makeatletter                    %allow @ in command names
\@addtoreset{equation}{section}  %reset equation count in each section
\makeatother                     %disallow @ in command names

\begin{document}

\title{Pathways toward understanding \\ Macroscopic Quantum Phenomena\footnote{Invited talk presented by BLH}}

\author{B L Hu and Y Suba\c{s}{\i}}
\address{Maryland Center for Fundamental Physics and Joint Quantum Institute \\ University of Maryland, College Park, Maryland 20742, USA}
\ead{blhu@umd.edu,ysubasi@umd.edu}

\begin{abstract} Macroscopic quantum phenomena refer to quantum
  features in objects of `large' sizes, systems with many components
  or degrees of freedom, organized in some ways where they can be
  identified as macroscopic objects.  This emerging field is ushered
  in by several categories of definitive experiments in
  superconductivity, electromechanical systems, Bose-Einstein
  condensates and others. Yet this new field which is rich in open
  issues at the foundation of quantum and statistical physics remains
  little explored theoretically (with the important exception of the
  work of  A J Leggett \cite{Leggett}, while touched upon or implied
  by several groups of authors represented in this conference. Our
  attitude differs in that we believe in the full validity of quantum
  mechanics stretching from the testable micro to meso scales, with no
  need for the introduction of new laws of physics.) This talk
  summarizes our thoughts in attempting a systematic investigation into some key foundational issues of quantum macroscopic phenomena, with the goal of ultimately revealing or building a viable theoretical framework. Three major themes discussed in three intended essays are the large N expansion \cite{MQP1}, the correlation hierarchy \cite{MQP2} and quantum entanglement \cite{MQP3}. We give a sketch of the first two themes and then discuss several key issues in the consideration of macro and quantum, namely, a) recognition that there exist many \textit{levels of structure }in a composite body and only by judicious choice of an appropriate set of \textit{collective variables} can one give the best description of the dynamics of a specific level of structure.  Capturing the quantum features of a macroscopic object  is greatly facilitated by the existence and functioning of these collective variables; b) \textit{quantum entanglement}, an exclusively quantum feature \cite{Schrodinger}, is known to persist to high temperatures \cite{VedralRMP} and large scales \cite{LutzChain} under certain conditions, and may actually decrease with increased connectivity in a quantum network \cite{EntQnet}. We use entanglement as a measure of quantumness here and pick out these somewhat counter-intuitive examples to show that there are blind spots worthy of our attention and issues which we need to analyze closer. Our purpose is to try to remove the stigma that quantum only pertains to micro, in order to make way for deeper probes into the conditions whereby quantum features of macroscopic systems manifest.
\end{abstract}

%\vskip .5cm
%\centerline{\small {\it-- Invited talk presented by BLH at the DICE 2012 meeting.  Proceedings to appear in J. Physics (Conf. Series)}}

\vskip .5cm

Macroscopic quantum phenomena (MQP) manifest in a number of systems,
superconductivity is probably the oldest often referred to, Bose-Einstein condensate (BEC)
\cite{MQPBEC} and electro- and opto-mechanical devices \cite{MQPnem,MQPom} are amongst the recent excitements. It is a relatively new research venue, with exciting
ongoing experiments and bright prospects, yet with surprisingly little
theoretical activity. From the traditional view that macroscopic objects are classical and quantum describes the microscopic realm, MQP appears like a transgression.  This of course is what makes it interesting intellectually.
This simplistic and hitherto rarely challenged view needs to be scrutinized anew, perhaps eventually with much of the
conventional wisdoms repealed. In a series of papers in preparation we attempt to explore  systematically into some key foundational issues of MQP, with the hope of finding  a viable theoretical framework for this new endeavour. The three major themes discussed in these three essays are the large N expansion, the correlation hierarchy and quantum
entanglement for systems of ``large'' sizes, with many components or degrees of freedom. Before delving into the subject proper, to orient ourselves,  we excerpt the beginning of \cite{MQP1} which charts the interloping and intersecting domains of quantum /classical and micro/ Macro.

\section{Quantum / classical, micro / macro}
\label{sec:Q/C,M/M}

\vskip .5cm
There are many ways to deal with the issue of quantum-classical
correspondence \cite{QCCDrexel}. In the most common and traditional
view the classical limit corresponds to  $\hbar \rightarrow 0$, or,
invoking the Bohr correspondence principle, the principal quantum
number of a system $n \rightarrow \infty $, or regarding the coherent
state as the ``most classical'' quantum state, or the Wigner function
as the ``closest to classical'' distribution. Less precise criteria also abound, such as the loose concept that a system at high temperature behaves classically, or viewing the thermodynamic / hydrodynamic limits (of a quantum system) as classical. (For a description of the various criteria, see, e.g., \cite{HuZhaUnc}). There are holes in almost all of the above common beliefs.  A more sophisticated viewpoint invokes decoherence, the process whereby a quantum system loses its coherence (measured by its quantum phase information) through interaction with its environment \cite{envdec}.   In this work we examine an alternative perspective, as the folklore goes,  that quantum pertains to the small (mass, scale) while classical to the large (size, multiplicity).  This common belief now requires a much closer scrutiny in the face of new challenges from macroscopic quantum phenomena (MQP), viz, quantum features may show up even at macroscopic scales. A common example is superconductivity where the Cooper pairs can extend to very large scales compared to interatomic distances and Bose-Einstein condensate (BEC) where a large number N of atoms occupy the same quantum state, the N-body ground state. Other examples include nanoelectromechanical devices \cite{nem} where the center of mass of a macroscopic classical object, the cantilever, obeys a quantum mechanical equation of motion. Experiments to demonstrate the quantum features such as the existence of interference between two macroscopic objects have been carried out, e.g., for $C^{60}$ molecules passing through two slits \cite{Arndt} or proposed mirror superposition experiments \cite{Marshall,EntSQL}.

A most direct account of the  difference between the microscopic and the macroscopic behaviours of a quantum system is by examining  N, the number of physically relevant (e.g. for atomic systems, forgetting about the tighter-bound substructures) quantum particles or components in a macroscopic object.  One may ask:  At what number of N will it be suitable to describe the object as mesoscopic with  qualitatively distinct features from microscopic and macroscopic? For classical systems  significant advances  in the recent decade have been made in providing a  molecular dynamics basis to the foundations of thermodynamics  \cite{Dorfman}, relating the macroscopic thermodynamic behaviour of a gas to the chaotic dynamics of its molecular constituents.  One could even calculate the range in the number of molecules where a microscopic system begins to acquire macroscopic behaviour and hence identify the approximate boundaries of mesoscopia \cite{Gaspard}.  For quantum systems one needs to deal with additional concerns of quantum coherence and entanglement which are critically important issues in quantum information processing (QIP) \cite{QIPsi}. A fundamental issue in QIP is how the performance of a quantum information processor alters as one scales the system up. This dependence on N  is known as the ``scaling" problem \cite{DVcriteria}.

There are many important and interesting issues of MQP,  one subset of
special interest to us is how quantum expresses itself in the
macroscopic domain since usually macro conjures classicality. Thus
even the simplest yet far from naive questions need to be reconsidered
properly. For example, why is it that an ostensibly macroscopic object
such as a cantilever should follow a quantum equation of motion.  This
``center-of-mass axiom'' is implicitly assumed in many descriptions of MQP but rarely justified. The conditions upon which this can be justified are explored in \cite{CHY} with the derivation of a master equation for N coupled harmonic oscillators (NHO) in a finite temperature harmonic oscillator bath, in the form of the HPZ master equation \cite{HPZ}, extending earlier work for 2HO \cite{ChouYuHu}. (A mathematically more vigorous and complete treatment of NHO system is given in \cite{FlemingNHO}.)  Presently we are continuing to explore the conditions where one could infer macroscopic quantum behavior, specifically in terms of the existence and degree of quantum entanglement in this coupled NHO model. One aspect is in terms of entanglement at finite temperature
\cite{AEPW02,Anders,AndWin,ZHH} and large distance \cite{LutzChain,SZH},  the other in terms of entanglement between different levels of structure (micro to meso to macro) \cite{EntMm,QTDNHO}  and the crucial role in a judicious choice of the appropriate collective variables \cite{Martins,MQP3}. This is discussed in Section 3.
In Section 4 we use the results of a recent paper on complex  quantum network \cite{EntQnet} to illustrate the somewhat counterintuitive finding that entanglement does not necessarily increase with connectivity but varies with the strength of coupling and the type of connectivity. (See also \cite{MacXmeso}.) Our understanding of this aspect will be described in a future work \cite{SHqNet}.

\section{Pathways toward understanding macroscopic quantum phenomena}

\vskip .5cm

In what follows we present two pathways as explored in two recent essays \cite{MQP1,MQP2}, the first concerns what macroscopic means -- large size? number? What about the degree of complexity of its constituents? What if the constituents are non-interacting versus interacting? Weakly interacting versus strongly interacting? The second pathway explores how quantum correlations and fluctuations impact on MQP using the n-particle-irreducible (nPI) representation.

\subsection{Pathway 1: From the large N perspective}

\vskip .2cm

In this paper we use different theories in a variety of contexts to
examine the conditions or criteria whereby a macroscopic quantum
system may take on classical attributes, and, more interestingly, that
it keeps some of its quantum features. The theories we consider here
are, the $O(N)$ quantum mechanical model, semiclassical stochastic
gravity and gauge / string theories; the contexts include that of a
``quantum roll'' in inflationary cosmology,  entropy generation in
quantum Vlasov equation for plasmas, the leading order and
next-to-leading order large N behaviour, and hydrodynamic /
thermodynamic limits. The criteria for classicality  in our
consideration include the use of uncertainty relations, the
correlation between classical canonical variables, randomization of
quantum phase, environment-induced decoherence, decoherent histories of hydrodynamic variables,  etc. All this exercise is to ask only one simple question: Is it really so surprising that quantum features can appear in macroscopic objects? By examining different representative systems where detailed theoretical analysis has been carried out, we find that there is no a priori good reason why quantum phenomena in macroscopic objects cannot exist.

\subsection{Pathway 2: From the correlation, coupling and criticality perspectives}
\vskip .2cm
In this sequel paper we explore how  macroscopic quantum phenomena  can be measured or understood from the behavior of quantum correlations which exist in a quantum system of many particles or components and how the interaction strengths change with energy or scale, under ordinary situations and when the system is near its critical point.
We use the nPI (master) effective action related to the Boltzmann-BBGKY / Schwinger-Dyson hierarchy of equations as a tool for systemizing the contributions  of higher order correlation functions to the dynamics of lower order correlation functions. Together with the large $N$ expansion discussed in our first paper \cite{MQP1} we explore 1) the conditions whereby an H-theorem is obtained, which can be viewed as a signifier of the emergence of macroscopic behavior in the system.  We give two more examples from past work: 2) the nonequilibrium dynamics of $N$ atoms in an optical lattice under the large $\cal N$ (field components), $2PI$ and second order perturbative expansions, illustrating how $N$ and $\cal N$ enter in these  three aspects of quantum correlations, coherence and coupling strength. 3) the behavior of an interacting quantum system near its critical point, the effects of quantum and thermal fluctuations and the conditions under which the system manifests infrared dimensional reduction. 
We also discuss how the effective field theory concept bears on
macroscopic quantum phenomena: the running of the coupling parameters
with energy or scale imparts a dynamical-dependent  and an
interaction-sensitive definition of ``macroscopia''.

\section{Levels of structure consideration in the micro-macro divide}

\vskip .5cm
Before we describe the third pathway toward understanding MQP, namely, using quantum entanglement in a system as a measure of its ``quantumness", it is useful to ponder on a related issue we explored a bit along the first pathway, namely,  what exactly is ``largeness" which usually is connected with macroscopic? Do all the basic constituents of a large object contribute equally towards its quantum feature? (This point is highlighted in footnote 2 of \cite{CHY}.) We actually know how these basic constituents are and how they are organized. A $C^{60}$ molecule is made of carbon atoms, each atom is made of nuclei and electrons, each nucleus contains a certain number of protons and neutrons, each of them in turn is made up of quarks and gluons. Are we to simply count the number of quarks /gluons or protons /neutrons when we say an object is macroscopic? Obviously the tight binding of them to form a nucleus enters into our consideration when we treat the nucleus as a unit which maintains its own more or less distinct features and dynamics. Thus when one talks about the mesoscopic or macroscopic behavior of an object one needs to specify which level of structure is of special interest and how important each level contributes to these characteristics. The coupling strength between constituents at each level of structure (e.g., inter-atomic) compared to that structure's
coupling with the adjacent and remaining levels (which can be treated as an environment to this specific level of structure in an effective theory description, and its influence on it represented as noise  \cite{CalHuEFT}) will determine the relative weight of each level of structure's partaking of the macroscopic object's overall quantum behavior. The best description of the behavior and dynamics of a particular level of structure is given by an effective theory for the judiciously chosen ``collective variables". Nuclear forces in terms of QCD is an illustrative example.

\subsection{Choose the right collective variables before considering their quantum behavior}

\vskip .2cm
Same consideration should enter when one looks for the ``quantumness"
of an object, be it of meso or macro scale. One can quantize any
linear system of whatever size, even macroscopic objects,  such as
sound waves from their vibrations. Giving it a name which ends with an
``on'' such as phonon and crowning it into a quantum variable is almost frivolous compared to the task of identifying the correct level of structure and finding the underlying constituents, the atoms in a lattice in this example, and their interactions.   
Constructing the relevant collective variables which best capture the salient physics of interest should come before one considers their quantum features. Thus, viewed in this perspective in terms of collective variables,  we see that  quantum features need not be restricted to microscopic objects. In fact `micro' is ultimately also a relative concept as new  ``elementary" particles are discovered which make up the once regarded `micro' objects.

We illustrate this idea with two examples below, one on the relevance of the center of mass variable in capturing the quantum features of a macroscopic object, the other on the description of entanglement between two macroscopic objects.

\subsubsection{The quantum and macroscopic significance of center of mass variable.}

\vskip .2cm
We can ask the question: what are  the conditions upon which the mechanical and statistical mechanical properties of a macroscopic object can be described adequately in terms mainly of its center-of-mass (COM) variable kinematics and dynamics as captured by a master equation (for the reduced density matrix with the environmental variables integrated out).
This is an implicit assumption made in many MQP investigations, namely, that the quantum mechanical behavior of a macroscopic object, like the nanoeletromechanical oscillator \cite{MQPnem,nem}, mirror \cite{Marshall}, or a $C^{60}$ molecule \cite{Arndt}, placed in interaction with an environment -- behavior such as quantum decoherence, fluctuations, dissipation and entanglement--  can be captured adequately by its COM behavior. For convenience we refer to this as the ``COM axiom''. This assertion is intuitively reasonable, as one might expect it to be true from normal- mode decompositions familiar in classical mechanics, but when  particles (modeled by NHO) interact with each other (such as in a quantum bound state problem) in addition to interacting with their common environment, all expressed in terms of the reduced density matrix, it is not such a clear-cut result. At least we have not seen a proof of it.

With the aim of assessing the validity of the COM axiom the authors of \cite{CHY}  considered a system modeled by $N$ harmonic oscillators  interacting with an environment consisting of $n$ harmonic oscillators  and derived an exact non-Markovian master equation such a system in a bath with arbitrary spectral density and temperature. The authors outlined a procedure to find a canonical transformation to transform from the individual coordinates $(x_i, p_i )$ to the collective coordinates $(\tilde{X}_i, \tilde{P}_i ), i=1,...,N$ where $\tilde{X}_1, \tilde{P}_1$ are the center-of-mass coordinate and momentum respectively. In fact they considered a more general type of coupling between the system and the environment in the form $f(x_i)q_j$ (instead of the ordinarily assumed $x_i q_j$) and examined if the COM variable dynamics separates from the reduced variable dynamics. They noted that if the function $f(x)$ has the
property $\sum_{i=1}^{N} f(x_i) = \tilde{f}(\tilde{X}_1) + g(\tilde{X}_2,...,\tilde{X}_N)$, for example $f(x) = x$ or $f(x) = x^2$, one can split the coupling between the system and environment into couplings containing the COM coordinate and the relative coordinates. Tracing out the environmental degrees of freedom $q_i$, one can easily get the influence action which characterizes the effect of the environment on the system.

However, as the authors of \cite{CHY} emphasized, the coarse-graining made by tracing out the environmental
variables $q_i$ does not necessarily lead to the separation of the COM and the relative variables in the effective action. When they are mixed up and can no longer be written as the sum of these  two contributions, the form of the master equation will be radically altered as it would contain both the relative variable and the center-of-mass variable dynamics.

With these findings they conclude that for the N harmonic oscillators quantum Brownian model, the coupling between the system and the environment need be bi-linear, in the form $x_i q_j$,  for this axiom to hold. They also proved that the potential $V_{ij}(x_i - x_j)$ is independent of the center-of-mass coordinate.  In that case, one can say that the quantum evolution of a macroscopic object in a general environment is completely described by the dynamics of the center-of-mass canonical variables ($\tilde{X}_1,\tilde{P}_1$) obeying a master equation of the HPZ \cite{HPZ} type.

What is the relevance of this finding to MQP?  Within the limitation of the N harmonic oscillator model it conveys at least two points: 1) For certain types of coupling the center of mass (COM) variable of an object composed of a large number of constituents does play a role in capturing the collective behavior of this object  2) Otherwise, more generally, the environment-induced quantum statistical properties of the system such as decoherence and entanglement could be more complicated. (For a similar conclusion considering the cross level (of structure) coarse-graining, see \cite{EntMm}.)

We now look at the quantum entanglement between two macroscopic objects, comparing the entanglement between the micro-variables of their constituents in two types of couplings, one to one and one to many. The very different nature between these two types serves to illustrate the relevance of how the micro-constituents organize into a macro object and how  entanglement between collective variables reveals the quantum features of a macroscopic entity. This issue is raised by Martins \cite{Martins} in the consideration of the entanglement between two wavepackets each containing sub-levels.  

\subsection{M-M, m-m and m-M entanglement}

\vskip .2cm

We now apply the methods developed in \cite{CHY} to the study of
the entanglement between the COM variables of two macroscopic objects.
Each macroscopic object is modeled by $N$ identical harmonic
oscillators (NHO). However, unlike \cite{CHY}, we do not include an environment in our discussion because
our focus is on the entanglement between the two objects induced by
various types of direct interactions between their microscopic constituents. We denote
the coordinates and the momenta of the microscopic constituents of the first and second
macroscopic object by $\left\{ \xo{i},\po{i} \right\}$ and $\left\{
\xt{i} , \ptw{i} \right\}$ respectively. The interactions between the
microscopic constituents of one macroscopic object are assumed to be
functions of the difference of variables only and we restrict
ourselves to bilinear couplings between the microscopic constituents
of the two macroscopic objects. The total Hamiltonian is then given
by:
\begin{eqnarray}
  H_1 &=& \sum_{i=1}^{N} \left( \frac{\po{i}^2}{2 M}
  +\frac{1}{2} M \om^2 \xo{i}^2 \right) + \sum_{i\ne j}^N
  \Vo{i}{j} ,\\
  H_2 &=& \sum_{i=1}^{N} \left( \frac{\ptw{i}^2}{2 M}
  +\frac{1}{2} M \om^2 \xt{i}^2 \right) + \sum_{i\ne j}^N
  \Vt{i}{j} ,\\
  H_I &=& \sum_{i, j}^N G_{ij} \xo{i} \xt{j} .
\end{eqnarray}
The canonical transformation described in the Appendix A. of
\cite{CHY} can be applied to each object
separately to yield a new set of phase space variables $\left\{
\Xo{i}, \Po{i}
\right\}$
and $\left\{
\Xt{i}, \Pt{i}
\right\}$
and the associated masses $\Mo{i}$. 
Here $\Xo{1} = \frac{1}{N}\sum_{n=1}^{N} \xo{i}$ and
$\Xt{1}= \frac{1}{N}\sum_{n=1}^{N} \xt{i}$ are the COM variables. The Hamiltonians of the macroscopic
objects can be written in terms of these variables as:
\begin{eqnarray}
  H_1 &= \sum_{i=1}^{N} \left( \frac{\Po{i}^2}{2 \Mo{i}}
  +\frac{1}{2} \Mo{i} \om^2 \Xo{i}^2 \right) + \tilde{V} \left(
  \Xo{2}, \cdots , \Xo{N}
  \right) = H_{1COM} + H_{1REL} ,\\
  H_2 &= \sum_{i=1}^{N} \left( \frac{\Pt{i}^2}{2 \Mt{i}}
  +\frac{1}{2} \Mt{i} \om^2 \Xt{i}^2 \right) + \tilde{V} \left(
  \Xt{2}, \cdots , \Xt{N}
  \right) = H_{2COM} + H_{2REL} .
\end{eqnarray}
It has been proven in \cite{CHY} that the potential $\tilde{V}$ is not a
function of the COM variable. This is a consequence of the form
assumed for
the potential energy.
For a general bilinear coupling characterized by $G_{ij}$ the interaction Hamiltonian $H_I$ can take a complicated form,
possibly mixing the COM variable with the relative variables.
Inspired by Martins\cite{Martins}, in
what follows we will focus on two particular choices of $G_{ij}$. The use of the new set of canonical variables which
include the COM will help interpret the behaviour of macroscopic
entanglement.

\subsubsection{Pairwise interaction pattern.}

The pairwise interaction pattern is defined by $G_{ij} = G
\delta_{ij}$. In other words one oscillator from object one couples to
one oscillator from object two, and all such couplings have the same
strength.
Using the canonical transformation of \cite{CHY} it can be shown that the
interaction Hamiltonian takes the form:
\begin{eqnarray}
  H_I &= \sum_{i}^{N} \frac{G}{M} \Mo{i} \Xo{i} \Xt{i} .
\end{eqnarray}
Note that pairwise interactions among the original variables
translate into pairwise interactions among the transformed
variables. A very important difference is that whereas the pairwise
interactions in the original variables were all equal strength, the
strength of the interactions scale with the mass of the
variables after the transformation. As a result the
relative strength of interactions between variable pairs are the
same for all the variables, including the COM. To see this
explicitly let us consider the case with $\tilde{V}=0$
for simplicity,
namely the micro-constituents of each macroscopic object do not
interact with each other. Then we rescale the coordinates by
$\Xo{i}\rightarrow \Xo{i} \sqrt{M/\Mo{i}}$ and $\Xt{i}\rightarrow
\Xt{i} \sqrt{M/\Mt{i}}$, after which the Hamiltonian takes the form:
\begin{eqnarray} 
  H &= \sum_{i=1}^{N} \left( \frac{\Po{i}^2}{2 M} + \frac{1}{2}M \om^2
  \Xo{i}^2 +
  \frac{\Pt{i}^2}{2 M} + \frac{1}{2}M \om^2 \Xt{i}^2 + G \Xo{i} \Xt{i}
  \right) .
\end{eqnarray}
In this form it is easy to see that the effective strength of interactions in
the COM variable is the same as the effective strength of interactions in all
the other variables. Hence the pairwise interaction pattern will
induce the same amount of entanglement between pairs of transformed
variables, without distinguishing the COM variable. Entanglement
between non-COM variables
would be effected if the interactions among the oscillators within the
same object, i.e. $V_{ij}$, are not set to zero.

If we only focus on the effect of
the pairwise interactions, it is fair to say that such interactions
couple the pairwise transformed variables with equal effective
strength independent of the size $N$ of the macroscopic objects.
As a consequence we expect the behavior of entanglement between the corresponding variables of the objects to be independent of the size of the macroscopic objects, even for the COM variable, as well. For instance, at a given temperature the amount of entanglement between the two corresponding variables of the objects will not depend on
$N$. The critical  temperature above which the entanglement ceases to exist also should not depend on $N$.

\subsubsection{One-to-all interaction pattern.}

The one-to-all interaction pattern is characterised by $G_{ij}=G$.
Then it is easy to see that the interaction Hamiltonian in the
transformed variables takes the form:
\begin{eqnarray}
  H_I &= N^2 G \Xo{1} \Xt{1} .
\end{eqnarray}
Note that \ota interaction pattern corresponds to a coupling only between
the COM variables of the macroscopic objects, the relative variable
Hamiltonian is unaffected.
Thus \ota pattern differs from the pairwise pattern in that it
distinguishes the COM variable. Moreover if we perform the same
rescaling of the previous section to determine the effective strength
of this coupling we get:
\begin{eqnarray} 
  H_{COM} = \frac{\Po{1}^2}{2 M} + \frac{1}{2}M \om^2 \Xo{1}^2 +
  \frac{\Pt{1}^2}{2 M} + \frac{1}{2}M \om^2 \Xt{1}^2 + N G \Xo{i}
  \Xt{i} .
\end{eqnarray}
We see that the effective strength of the coupling increases with
increasing $N$ for the \ota pattern. Thus in this case we expect the
entanglement between the COM variables to increase with increasing size of the macroscopic
objects and survive at higher temperatures. The \ota interaction
pattern is crucial for the scaling of the entanglement of COM variables with $N$. Hence it is important to
investigate if this type of interaction pattern can occur in realistic
situations and if so how common it is.

\section{Macroscopic quantum phenomena from the entanglement perspective}

\vskip .5cm

Entanglement is considered as a uniquely quantum feature \cite{Schrodinger}, and quantum is habitually viewed as
a zero or low temperature phenomenon or pertaining only to small systems.  Both of these conditions are now being reconsidered, the ``small system" restriction facing new challenges from macroscopic quantum
phenomena (MQP) we are considering here.  The ``low temperature" restriction is lifted by theoretical observations \cite{VedralRMP,WVB08} and experimental proposals \cite{GalvePRL} that entanglement can survive at high temperatures,  some even speculate that it is witnessed in biological systems \cite{QEntBio}.

\subsection{Quantum entanglement at high temperatures and long distances?}

\vskip .2cm
Theoretical analysis of this issue for such systems has been carried out for coupled oscillator chain (1D) \cite{AEPW02} or lattices (2D or 3D), where bounds and phase diagrams showing entangled and separable states were obtained by Anders \cite{Anders}. For a nice expos\`e of the general issues on this topic we refer readers to the Discussion and Conclusion Section of \cite{AndWin}.

\textit{Thermal or hot entanglement} can be studied by generalizing the (zero-temperature) quantum field mimicking a harmonic lattice to a thermal (finite temperature) field. In terms of model description
quantum entanglement between two inertial  harmonic oscillators interacting via a zero-temperature quantum field was  studied earlier by Lin and Hu  \cite{LinHu09} who showed that in addition to the temporal evolution of their entanglement there is also a dependence on their spatial separation.
This generalization is done recently in \cite{ZHH} wherein both the temporal and spatial dependence of quantum entanglement studied before are shown to be sensitive to temperature variation. These authors also considered thermal entanglement in a harmonic lattice but with strong coupling, extending the comprehensive study of \cite{AndWin}.

Another aspect is how much quantum entanglement can survive at large distance. \textit{Long range entanglement} in a coupled oscillator chain was  claimed by Wolf et al \cite{LutzChain}. Their setup of two harmonic oscillators interacting with a one dimensional harmonic lattice in a Gibbs state and their choice of parameters (continuum limit) map snugly to the thermal field model mentioned above. There are advantages in approaching the thermal entanglement issue for continuum systems from a quantum field theory perspective. Besides the technical ease to perform integrals over finite sums, the special properties of lower-dimensional systems (such as the Coleman-Mermin-Wegner theorem and the Berezinskii-Kosterlitz-Thouless phase transition) are well known, in some cases aided by elegant conformal field theory properties. This calculation is presently carried out in \cite{SZH} where existence of zeros in the spectral density is found to be the cause of long range entanglement. How general is this tie has yet to be decided.

\subsection{Quantum networks: more connected not always more entangled}

\vskip .2cm

Finally we mention the results from a recent paper to illustrate a point on the relation between connectivity in a quantum network and entanglement. The following are excerpted from \cite{EntQnet}.

A network is defined as a set of $N$ nodes and $E$ edges accounting for their pairwise interactions. The
network is usually characterized by its adjacency matrix, $A$, with elements
$A_{ij}= 1$ if an edge connects nodes $i$ and $j$ while $A_{ij}=0$ otherwise.
We restrict attention to the undirected network where $A_{ij}= A_{ji}$. The Laplacian is related to the adjacency matrix by $L_{ij}= k_i \delta_{ij} - A_{ij}$, where $k_i = \sum_j A_{ij}$ is the connectivity of node $i$, {\it i.e.}, the number of nodes connected to $i$.

We can represent the nodes of the network by identical quantum oscillators
interacting in accordance to the network topology encoded in $L$. The Hamiltonian
of the harmonic quantum network is given by:
\begin{equation}
\label{Hnet}
H_{\rm network} =\frac{1}{2} \Big (   {\bf p}^{\rm T} {\bf p} + {\bf x}^{\rm T}
(\mathbb{I} + 2cL) {\bf  x}\Big )\, ,
\end{equation}
here $\mathbb{I}$ is the $N \times N$ identity matrix, $c$ is the coupling strength between
connected oscillators while  ${\bf p}^{\rm T}= (p_1,  p_2,..., p_N)$ and ${\bf
x}^{\rm T}= (x_1, , x_2,...,x_N)$ are the operators corresponding to the momenta
and positions of nodes respectively, satisfying the usual commutation relations:
$[{\bf x}, {\bf p}^{\rm T}] = {\mbox i}\hbar\,\mathbb{I}$.

The properties of the ground state of Hamiltonian (\ref{Hnet}) can be studied to quantify
the amount of information each element of a network shares with the rest of the system
via quantum fluctuations. Even at zero temperature the nodes are not at rest due to Heisenberg uncertainty
principle. Their spatial fluctuations depends on the pattern of physical interactions, {\em i.e.}, the network
structure. To show this, the authors of \cite{EntQnet} consider the partition of the network
into a node, say $i$, and its complement $i^{\rm c}$, {\em i.e.} the rest of the
network.
The mutual information shared by the two parties is  given by:
\begin{equation}
\label{mi}
{\mathcal{I}} (i | i^{\rm c} )= S_i + S_{i^{\rm c}} - S_{\rm tot}\;.
\end{equation}
Here $S_i$  and $S_{i^{\rm c}}$ are marginal entropies and $S_{\rm tot}$ is the
total entropy of the network. It is natural to choose the Von Neumann entropy to
quantify the quantum information of the system, yielding $S_{\rm tot} = 0$ for the ground state (as it is a {\em pure state}).
Since the total network
is in its ground (and pure) state we have $S_i= S_{i^{\rm c}}=\mathcal{I}(i|i^{\rm c})/2$. Therefore, the information that a node shares
with the network is intrinsically due to quantum correlations. Equivalently, the
mutual information is, itself, a measure of the entanglement (quantified by $S_i$)
between a single node and the rest of the system.

The authors then quantify the entanglement entropies of nodes embedded in
different network topologies. They consider two homogeneous network
substrates: {\em (i)} Random Regular Graphs (RRG), in which all the nodes have
the same number of contacts ($k_i=\langle k\rangle, \forall i$), and {\em (ii)}
Erd\H{o}s-R\'enyi (ER) networks \cite{ER}, for which the probability of finding a
node with $k$ neighbors, $P(k)$, follows a Poisson distribution so that most of
the nodes have a degree $k$ close to the average $\langle k\rangle$.
They also analyze two networks having a scale-free (SF) pattern for the probability
distribution, $P(k)\sim k^{-3}$, constructed by means of a configurational
random model (SF-CONF) \cite{conf}  and the Barab\'asi-Albert model (SF-BA)
\cite{BA}. Their results are presented in plots of the average entanglement entropy of a node
with connectivity $k$, i.e $\langle S_k\rangle$, vs $k$ for the three network models: ER, SF-CONF and SF-BA.

Interesting features can be gleaned from the figures in  \cite{EntQnet}:
Fig.2 plots $\ask$ for fixed average connectivity $\langle k
\rangle$ and 4 different values of coupling strength $c$.  It shows that the average entanglement of a node with given connectivity $k$ increases with increasing coupling strength. As a check the case $c=0$ corresponds to non-interacting oscillators which in their ground state are not entangled. It is expected that as the interactions get stronger the ground state becomes more and more entangled.
Fig.3 plots $\ask$ vs $k$ for fixed $c$ and different values of $\ak$.
It can be seen that for fixed $k$ the entanglement $\ask$ increases
for decreasing $\ak$ for all graphs.

Here we offer some tentative explanations on such qualitative behaviors. 
We can understand this using the idea of monogamy of entanglement, which says that a system which is fully
entangled to another system cannot be entangled to a third system.
Keeping $k$ fixed while decreasing $\ak$ amounts to reducing the
connections the neighbors of the node of interest has. Thus
its neighbors have less neighbors to get entangled with. As a result
they can be entangled more with the node of interest.

Another observation we can make from both Fig.2 and Fig.3, which is less intuitive, is the fact that
$\ask$ flattens out for ER for large $\ak$ and first rises and then
falls for SF. This indicates that for ER the nodes with large connectivity have all the same amount of entanglement with the rest of the network. On the other hand for SF there is an optimal
number for the connectivity such that those nodes with the optimal
number of connections have the highest amount of entanglement with the
rest of the network.

How can we make sense of this? Naively one expects the entanglement to increase with increasing
number of connections, because more connections means more
correlations. However entanglement is not just correlations. There may
be a competition between correlations and monogamy of entanglement (or
some argument using properties of quantum mutual information)
that causes the rise and fall of entanglement in SF and the saturation
in ER.  Our own investigation into the entanglement behavior of quantum oscillator networks is under way \cite{SHqNet}.

\section{Conclusion}
\vskip .5cm

In this talk we presented several pathways toward understanding MQP,
identified some key issues we need to address or be concerned with and
provided some examples to illustrate possibly counter-intuitive
behavior. For the present issue of quantum entanglement, specifically,
using its extent and behavior to measure the quantumness of a system,
we pointed out the necessity to recognize the levels of structure and
the usefulness of collective variables in describing a macroscopic
composite object when we try to identify its quantum features. One
needs to be aware of the qualitative differences between the
entanglement amongst the micro-constituents and that between
collective variables which reveal MQP. We mentioned  entanglement at
finite temperature and at long ranges, and  used quantum coupled
oscillator networks to illustrate the varying degrees of entanglement
with different types of connectivity. We hope with these sampling of
ideas, approaches and illustrative examples we can stimulate greater
interest in how to think about the quantum nature of macroscopic
objects, and,  perhaps along the way, gain a deeper understanding of quantum physics itself.

\vskip .5cm

\section* {Acknowledgments} BLH wishes to thank Professor Thomas Elze for his invitation to this rather unique series of meetings where issues at the foundations of gravitation and quantum and statistical mechanics are explored.  He is also grateful to Professor M. C. Chu for his gentle hospitality at the Chinese University of Hong Kong where this paper was finished. 

\vskip 1cm

\bibliography{DICE2013}{}
\bibliographystyle{iopart-num}

\end{document}